# Experimental observation of the gate-controlled reversal of the anomalous Hall effect in the intrinsic magnetic topological insulator MnBi$_2$Te$_4$ device

Shuai Zhang[1,*], Rui Wang[1,2,*], Xuepeng Wang[1,*], Boyuan Wei[1,*], Huaiqiang Wang[1], Gang Shi[3], Feng Wang[1], Bin Jia[1], Yiping Ouyang[1], Bo Chen[1], Qianqian Liu[1], Faji Xie[1], Fucong Fei[1], Minhao Zhang[1], Xuefeng Wang[4], Di Wu[1], Xiangang Wan[1], Fengqi Song[1,†], Haijun Zhang[1,†], Baigeng Wang[1,†]

[1] National Laboratory of Solid State Microstructures, Collaborative Innovation Center of Advanced Microstructures, and School of Physics, Nanjing University, Nanjing 210093, China.

[2] Department of Physics and Astronomy, Shanghai Jiao Tong University, Shanghai, 200240, China

[3] Beijing National Laboratory for Condensed Matter Physics, Institute of Physics, Chinese Academy of Sciences, Beijing 100190, China.

[4] National Laboratory of Solid State Microstructures, School of Electronic Science and Engineering and Collaborative Innovation Center of Advanced Microstructures, Nanjing University, Nanjing 210093, China

---

[*] These authors contribute equally.

[†] Corresponding authors. Email: S.F.(songfengqi@nju.edu.cn), H.Z.(zhanghj@nju.edu.cn). B.W.(bgwang@nju.edu.cn)

**Abstract**

Here we report the reserved anomalous Hall effect (AHE) in the 5-septuple-layer van der Waals device of the intrinsic magnetic topological insulator $MnBi_2Te_4$. By employing the top/bottom gate, a negative AHE loop gradually decreases to zero and changes to a reversed sign. The reversed AHE exhibits distinct coercive fields and temperature dependence from the previous AHE. It reaches the maximum inside the gap of the Dirac cone. The newly-seen reversed AHE is attributed to the competition of the intrinsic Berry curvature and the Dirac-gap enhanced extrinsic skew scattering. Its gate-controlled switching contributes a scheme for the topological spin field-effect transistors.

**Introduction**

The pursuit of high quality quantum anomalous Hall [1-5] devices ignited the upsurge of intrinsic magnetic topological insulators (IMTI) of $MnBi_2Te_4$ (MBT) [6-14] to avoid the possibly low strength of doped [15-17] or proximity-induced [18, 19] magnetism in previous magnetic topological insulators [20]. It was noticed with a surface band and predicted to be IMTI with a sizeable magnetic gap by several groups [8-10]. It is ferromagnetic in plane with out-of-plane antiferromagnetic interlayer coupling. Its thin film has been grown by Y. Gong *et al.*, which exhibits ferromagnetism-induced anomalous Hall loops [6]. Single antiferromagnetic crystals with the Neel temperature of about 23 K were grown by several groups and soon optimized to *n-p* transition by atomic doping [21]. The single crystals can be exfoliated to thin flakes, with odd number of layers presenting ferromagnetism and anomalous Hall responses. These efforts have led to the recent observation of quantized anomalous Hall plateau [22] up to 4.5 K and 6 T, as well as a transition from axion insulator to Chern insulator [23]. However, in the most intriguing zero-field regime, the device environment is made so complex by the coexistence of multiple physical interactions including antiferromagnetic, ferromagnetic, topological and spin-dependent scatterings, that the mining over this regime has proven hard. Here we fabricate the van der Waals (vdW) heterojunction (graphene/BN/MBT/$SiO_2$/Si), and focus on its electrical transport near zero field, where the interesting sign reversal of the anomalous Hall effect (AHE) appears. We find the gate-controlled switching between the positive and negative AHE loops, which makes the device a possible spin field effect transistor (FET). Its physics is then interpreted by the competition between the Berry curvature and Dirac-gap enhanced skew scatterings.

**Results**

**The negative AHE displaying the odd-layer ferromagnetism in MBT**

The MBT crystals, consisted by Te-Bi-Te-Mn-Te-Bi-Te septuple layers (SLs), are grown by the flux method [21]. With mechanical exfoliation we can obtain the MBT nanoflakes with thin thickness on the doped Si substrates coated with 300-nm $SiO_2$. An atomic force microscope graph for the 7.1-nm-thickness MBT nanoflake is shown in Fig. 1A, which means that it is 5 SLs, and would exhibit the ferromagnetism. We fabricate this sample by the standard lithography process. After that, we transfer the hexagonal boron nitride (h-BN) on the top of the MBT, followed by few-layers graphene being transferred on the h-BN, acting as the top gate ($V_{top}$), with the $SiO_2$/Si being the back gate ($V_{back}$). Then the dual-gate MBT field effect transistor device is obtained (Fig. 1B). The schematic of the measurement configuration is shown in the right panel of Fig. 1B.

The temperature ($T$) dependent longitudinal resistance ($R_{xx}$) without gate voltages is measured (Fig. 1C). The longitudinal resistance decreases first with the temperature decreasing, while when the temperature is below about 30 K, the resistance becomes increasing. At low temperatures, both the top gate and the back gate can modulate the resistance effectively. The dual-gate mapping at $T$ = 2 K is shown in Fig. 1D. Because of its thin thickness, the two topological surface states are coupled, and the two gates are dependent. We can find that the resistance increases with the gate voltages (both $V_{top}$ and $V_{back}$) decreasing, meaning that the carriers in our sample is $n$-type, and the Fermi level is above the charge neutral point (CNP). Also, most of the mapping is cyan, that's to say in these gate voltage regions, the resistance increases very slowly, indicating a high density of states. While in the lower left part of the mapping, the resistance varies greatly, which means that the Fermi level might be located in the topological surface states regimes.

We measure the magnetotransport with various gate voltages at $T = 2$ K. When the back-gate voltage is -40 V without applying the top-gate voltage, that is ($V_{top}$, $V_{back}$) = (0 V, -40 V), we can observe a clear AHE at $T = 2$ K (Fig. 1E). And the butterfly loop in $R_{xx}$ can also be observed, indicated by the arrows shown in the inset of the Fig. 1E. For clarity, we define this kind of AHE as negative AHE, because the $R_{yx}$ drops at the coercive field when the magnetic field sweeps from negative magnetic field to positive magnetic field (the red curve in Fig. 1E). This indicates the onset of the ferromagnetism in the 5-SLs device. Such ferromagnetism appears in all the devices with odd layer numbers (3, 5, 7, 9, 11 and 13) of SLs of MBT, while it becomes antiferromagnetism in the devices with even layer numbers of SLs. This is reasonable since the neighboring layers forms a pair of layers with opposite spin directions. The ordinary Hall effect (OHE) with a negative slope is consistent with the $n$-type carriers.

**Switching to the reversed AHE by the gate control**

Benefitting from the dual gates, we can tune the AHE in a large range of the

Fermi level. We measure the AHE with $V_{top}$ changing from 0 V to -16 V at fixed $V_{back}$ = -80 V first and focus on the AHE near zero fields (less than 2 T) and its sign. We find that with the decreasing $V_{top}$, the previous negative AHE becomes smaller, changes from negative to positive (Fig. 2A). Its loop reaches even larger than that at $V_{top}$ = 0. The transition point is at $V_{top}$= -8 V. Furthermore, when we change the $V_{back}$ from 0 V to -85 V at fixed $V_{top}$ = -12 V, the zero-field AHE also exhibits a similar sign reversal (Fig. 2B). In both processes, we find the Hall traces at high field keep the same direction, indicating the dominance of the same type of carriers. The previous negative zero-field AHE exhibits the same direction as the high-field Hall trace, also consistent with other work [22], while the positive AHE gives reversed direction. This means the built-in field of the IMTI has been switched nontrivially during the gate tuning, whose physics becomes the central point of this work below.

Detailed analysis confirms the distinct physics of the positive AHE. In Figs. 2D and 2E, we extract the value of the zero-field anomalous Hall resistance (AHR), denoted as $R_{AH}$, in order to distinguish from the high field AHR, and the focus of this work is the zero-field AHR. Here, the $R_{AH}$ exhibits sign reversal with the gate voltage decreasing. In the positive AHE regime, the $R_{AH}$ increases quickly up to over 0.4 kΩ; while in the negative AHE regime, the $R_{AH}$ varies gradually near 0.17 kΩ. The longitudinal resistance at zero field ($R_{xx0}$) corresponding to the $R_{AH}$ is also shown in the Figs. 2D and 2E. From the gate dependent $R_{xx0}$, as well as the dual-gate tuned resistance mapping and the slope of the OHE, we can know that the carrier is $n$-type in the gate voltage range, and the CNP is not achieved within the gate voltage range from (0 V, 0 V) to (-12 V, -80 V). The Fermi level is modulated to approaching the CNP by decreasing the gate voltage. Also, The $R_{xx0}$ increase monotonously with the gate voltage decreasing, and the highest value of $R_{xx0}$ can be up to about 10 kΩ, which should be contributed by the surface states [24]. Therefore, it means that the sign reversal is not related to the carrier type, but has a more profound implication.

Next, we fix the gate voltage as (-16 V, -80 V), and change a variety of temperatures to measure the temperature dependent AHE (Fig. 2C). The AHE would disappear when the temperature is over 17 K. But the sign of AHE is always positive, different from the temperature induced sign reversal [25], indicating the temperature is not the influence factor here. From the temperature dependent resistance, we can find the second-order differential of resistance to temperature exhibits a peak at 17.3 K, shown in the inset of Fig. 1C. It is consistent with the temperature that AHE disappears. It is possibly the Curie temperature of the devices. Figure 2F shows the zero-field AHR at different temperatures with fixed gate voltage as (-16 V, -80 V), as well as the longitudinal resistance at zero field. Both of them decrease with the temperature increasing.

**The scaling behavior of the reversed AHE**

To get insight into the sign reversal of the AHE, we extract the critical field from the first-order differential of $R_{yx}$ to $B$ (the inset of Fig. 3A), and plot the data in Figs. 3A and 3B. The coercive field ($\mu_0 H_{C1}$) at different gate voltage (Fig. 3A) exhibits different behavior when the AHE reverse its sign from negative to positive. In the negative AHE region, the $\mu_0 H_{C1}$ is almost unchanged at 0.46 T. But it drops clearly in the positive AHE region from 0.6 T to 0.48 T. And when the gate voltage is (-7 V, -80 V) or (-12 V, -70 V), there are two coercive fields, indicating the coexistence of both the negative and positive AHE. The coexistence regime is not our focus, which may have intriguing physics deserving further study. In Fig. 3B, the second critical field ($\mu_0 H_{C2}$) doesn't have a similar behavior with $\mu_0 H_{C1}$, and it varies very slowly around 2.1 T. It is consistent with high-field AHE, whose sign keeps unchanged in the whole gate voltage range. There seems being only a weak enhancement of $\mu_0 H_{C2}$ with the gate voltage decreasing.

In addition, from the longitudinal resistance, more aspects can be detected. The first-order differential of $R_{xx0}$ to the gate voltage exhibits a clear peak (Fig. 3C). And this peak corresponds to the transition point from negative AHE to positive AHE. We plot the two transition points in Fig. 1D as two black circles. A dash line connecting the two points indicates the peak of the first-order differential of $R_{xx}$ to the gate voltage. Therefore, the crossover of the AHE can be seen in the mapping of $R_{xx}$, and the green and red parts can be considered as the negative AHE and positive AHE, respectively. We can also calculate the sheet conductivity ($\sigma_{xx} = \rho_{xx}/(\rho_{yx}^2 + \rho_{xx}^2)$) and anomalous Hall conductivity (AHC, $\sigma_{xy} = \rho_{yx}/(\rho_{yx}^2 + \rho_{xx}^2)$), at different gate voltages, where $\rho_{xx} = R_{xx}(W/L)$ (W and L is the channel width and length) and $\rho_{yx} = R_{yx}$. In Fig. 3D, the relationship between the $\sigma_{xx}$ and zero-field AHC ($\sigma_{AH}$) is plotted, with a linear fitting (the red line).

In another 5 SLs MBT dual-gate device, we measure the magnetotransport in a larger range of gate voltages. In this device, the CNP is achieved. In Fig. 3E, we plot the $R_{AH}$ at various gate voltages. In this device, the positive $R_{AH}$ can reach a max value at fixed top gate while changing the back gate. The $R_{xx0}$ is plotted as a mapping in Fig. 3F. We can find that the $R_{xx0}$ and the $R_{AH}$ is maximal at the same gate voltages, indicated by the yellow dash line, which means that the positive $R_{AH}$ is largest at the CNP.

The similar sign-reversal of $\rho_{yx}$ was observed in several other materials [25-27]. For example, in topological insulator heterostructures $Cr_x(Bi_{1-y}Sb_y)_{2-x}Te_3$/ $(Bi_{1-y}Sb_y)_2Te_3$, a temperature-dependent sign reversal was attributed to the Rashba-induced splitting of bulk bands [25]; in magnetically doped topological insulators $(Bi,Mn)_2Se_3$, a hysteresis-free anomalous Hall effect with sign reversal was found and proposed as a result of two competing components from the system [27]. In comparison, here the sign reversal of AHE is observed in the intrinsic magnetic topological insulator, and shows a complete reversal of the hysteresis and is robust against varying temperatures up to the critical temperature $T_C$ = 17.3 K. This phenomenon has to be interpreted by some new model.

**The AHE reversal by the Dirac-gap enhanced spin skew scattering**

Here we propose a qualitative understanding of the reversal of hysteresis loop, which could be originated from competition between the Berry curvature and the skew scattering. From the change of the carrier density as well as the resistance (Fig. 1D and Figs. 2D and 2E), one can estimate that the chemical potential is tuned from the bottom of bulk band into the surface state with applying the gate, i.e., from (0 V, 0 V) to (-16 V, -80 V) (or (-12 V, -85 V)) in Figs. 2A and 2B.

Focusing on this energy range, we obtain the band structure of 5SL MBT thin film and the intrinsic AHC $\sigma_{xy}$ using the first-principles calculations, where the total magnetization is fixed as $M_Z > 0$. As shown, in the energy range investigated (marked by the dashed region in Fig. 4A), the contribution of the Hall conductance mainly comes from the surface state, which increases from a small negative value towards $-e^2/h$, where $e$ is the elementary charge and $h$ is the Planck constant, with the chemical potential tuning towards the surface states.

Then we further consider the effect of external magnetic field. Applying a positive field, $B > 0$, since the Berry curvature is not sensitive to the low field, $\sigma_{xy}$ will remain nearly unchanged until the field reaches $\mu_0 H_{C1}$ and induces the flip of total magnetization from $M_Z < 0$ to $M_Z > 0$. This in turn generates the reversal of Berry curvatures and therefore flips the Hall conductivity from $\sigma_{xy} > 0$ to $\sigma_{xy} < 0$. Similarly, the inverse effect takes place with applying a negative field $B < 0$. This leads to a hysteresis as shown by Fig. 4C.

With gradually applying the gate, since the Berry curvature increases monotonously as shown by Fig. 4A, the hysteresis will remain the same shape without sign reversal, if one only considers the contribution from the Berry curvature. It is well known that besides the Berry curvature, other extrinsic mechanisms, namely the skew scattering and side jump, can reshape the Hall conductance significantly [28, 29]. These effects could play important roles in 5SL MBT since there is always a net magnetization even at zero fields. From the scaling data in Fig. 3D, it is clear that $\sigma_{xy}$ is a linear function of $\sigma_{xx}$ with a constant shift, strongly indicating that the skew scattering effect has to be considered [28] in accompany with the intrinsic Berry curvature. Therefore, the total Hall conductance becomes a sum of two components, $\sigma_{xy}^{\text{tot}} = \sigma_{xy}^{\text{int}} + \sigma_{xy}^{\text{sk}}$. $\sigma_{xy}^{\text{sk}}$ is proportional to the total magnetization $M_Z(B)$ and therefore contributes to an opposite hysteresis as shown in Fig. 4C. Moreover, since $\sigma_{xy}^{\text{sk}}$ is proportional to scattering time $\tau$ which is inversely proportional to the density of states at the Fermi level, $\sigma_{xy}^{\text{sk}}$ will grow significantly when the chemical potential moves towards the surface band gap. Therefore, its effect becomes more and more manifested with applying the gate, and generates the reversal of the hysteresis of $\sigma_{xy}^{\text{tot}}$. It is worth to mention that, with increasing the external field, a full large mass gap at the surface state expects to be generated, which results in a significant $\sigma_{xy}^{\text{int}}$ at high field regime and dominates over $\sigma_{xy}^{\text{sk}}$. This would lead to a negative $\sigma_{xy}^{\text{tot}}$ at the large and positive field.

In summary, we study the AHE in the vdW heterojunction of an IMTI of MBT and find a sign reversal of the AHE controlled by the external gate. This reversal happens before the change of the carrier type and the reserved AHE loop reaches the maximum near the Dirac gap. Through the theoretical analysis, we propose that the sign reversal may originate from the competition between the intrinsic Berry curvature and the extrinsic skew scattering. This gate-controlled spin transport paves the way towards the topological spintronics and long-desired spin field effect transistors [30, 31].

**References**

[1] C.-Z. Chang, *et al.* Science **340**, 167 (2013).

[2] J. G. Checkelsky, R. Yoshimi, A. Tsukazaki, K. S. Takahashi, Y. Kozuka, J. Falson, M. Kawasaki and Y. Tokura. Nat. Phys. **10**, 731 (2014).

[3] C.-Z. Chang, W. Zhao, D. Y. Kim, H. Zhang, B. A. Assaf, D. Heiman, S.-C. Zhang, C. Liu, M. H. W. Chan and J. S. Moodera. Nat. Mater. **14**, 473 (2015).

[4] X. Kou, *et al.* Phys. Rev. Lett. **113**, 137201 (2014).

[5] A. J. Bestwick, E. J. Fox, X. Kou, L. Pan, K. L. Wang and D. Goldhaber-Gordon. Phys. Rev. Lett. **114**, 187201 (2015).

[6] Y. Gong, *et al.* arXiv:1809.07926 (2018).

[7] S. Huat Lee, *et al.* arXiv:1812.00339 (2018).

[8] J. Li, Y. Li, S. Du, Z. Wang, B.-L. Gu, S.-C. Zhang, K. He, W. Duan and Y. Xu. arXiv:1808.08608 (2018).

[9] M. M. Otrokov, *et al.* arXiv:1809.07389 (2018).

[10] D. Zhang, M. Shi, T. Zhu, D. Xing, H. Zhang and J. Wang. arXiv:1808.08014 (2018).

[11] M. M. Otrokov, I. P. Rusinov, M. Blanco-Rey, M. Hoffmann, A. Y. Vyazovskaya, S. V. Eremeev, A. Ernst, P. M. Echenique, A. Arnau and E. V. Chulkov. Phys. Rev. Lett. **122**, 107202 (2019).

[12] J. Q. Yan, Q. Zhang, T. Heitmann, Z. L. Huang, W. D. Wu, D. Vaknin, B. C. Sales and R. J. McQueeney. arXiv:1902.10110 (2019).

[13] J. Li, C. Wang, Z. Zhang, B.-L. Gu, W. Duan and Y. Xu. arXiv:1905.00642 (2019).

[14] R. C. Vidal, *et al.* arXiv:1903.11826 (2019).

[15] J. G. Checkelsky, J. Ye, Y. Onose, Y. Iwasa and Y. Tokura. Nat. Phys. **8**, 729 (2012).

[16] S.-Y. Xu, *et al.* Nat. Phys. **8**, 616 (2012).

[17] Y. L. Chen, *et al.* Science **329**, 659 (2010).

[18] F. Katmis, *et al.* Nature **533**, 513 (2016).

[19]  C. Tang, *et al.* Sci. Adv. **3**, e1700307 (2017).

[20]  Y. Tokura, K. Yasuda and A. Tsukazaki. Nat. Rev. Phys. **1**, 126-143 (2019).

[21]  B. Chen, *et al.* arXiv:1903.09934 (2019).

[22]  Y. Deng, Y. Yu, M. Zhu Shi, J. Wang, X. H. Chen and Y. Zhang. arXiv:1904.11468 (2019).

[23]  C. Liu, Y. Wang, H. Li, Y. Wu, Y. Li, J. Li, K. He, Y. Xu, J. Zhang and Y. Wang. arXiv:1905.00715 (2019).

[24]  Y. Xu, I. Miotkowski, C. Liu, J. Tian, H. Nam, N. Alidoust, J. Hu, C.-K. Shih, M. Z. Hasan and Y. P. Chen. Nat. Phys. **10**, 956 (2014).

[25]  K. Yasuda, R. Wakatsuki, T. Morimoto, R. Yoshimi, A. Tsukazaki, K. S. Takahashi, M. Ezawa, M. Kawasaki, N. Nagaosa and Y. Tokura. Nat. Phys. **12**, 555 (2016).

[26]  D. Chiba, A. Werpachowska, M. Endo, Y. Nishitani, F. Matsukura, T. Dietl and H. Ohno. Phys. Rev. Lett. **104**, 106601 (2010).

[27]  N. Liu, J. Teng and Y. Li. Nat. Commun. **9**, 1282 (2018).

[28]  N. Nagaosa, J. Sinova, S. Onoda, A. H. MacDonald and N. P. Ong. Rev. Mod. Phys. **82**, 1539-1592 (2010).

[29]  N. A. Sinitsyn. J. Phys.: Condens. Matter **20**, 023201 (2007).

[30]  I. Žutić, J. Fabian and S. Das Sarma. Rev. Mod. Phys. **76**, 323-410 (2004).

[31]  V. Baltz, A. Manchon, M. Tsoi, T. Moriyama, T. Ono and Y. Tserkovnyak. Rev. Mod. Phys. **90**, 015005 (2018).

**Acknowledgements**

The authors gratefully acknowledge the financial support of the National Key R&D Program of China (Grant No. 2017YFA0303203 and No. 2018YFA0306800), the National Natural Science Foundation of China (Grant No. U1732273, No. U1732159,

**Figure Captions:**

**Figure 1. The 5 SL MBT device and the onset of the reversed AHE.**

(A) The atomic force microscope graph of the 7.1-nm-thickness MBT nanoflake (5 SL). And the height of the line shows the thickness of the MBT nanoflake is about 7.1 nm. The white scale bar is 5 μm. (B) The optical graph and configuration of the 5 SL MBT device. The red scale bar is 20 μm. (C) Temperature dependent resistant. The inset is the second-order differential of resistance to temperature, which shows a peak at 17.3 K. (D) Dual gate tuned resistant at 2 K, where the dashed line marks the onset of reversed AHE loops. (E) The AHE at $T= 2$ K with $V_{back} = -40$ V. And the inset shows the butterfly loop.

**Figure 2. The AHE sign reversal and its temperature dependence.**

(A) The AHE at different $V_{top}$ when $V_{back} = -80$ V and $T = 2$ K. The sign reverses with the $V_{top}$ decreasing. (B) The AHE at different $V_{back}$ when $V_{top} = -12$ V and $T = 2$ K. The sign reverses with the $V_{back}$ decreasing. (C) The AHE at different temperatures when $V_{top} = -16$ V and $V_{back} = -80$ V. (D) The $R_{AH}$ and $R_{xx0}$ at different topgate voltage when $V_{back} = -80$ V. (E) The $R_{AH}$ and $R_{xx0}$ at different backgate voltage when $V_{top} = -12$ V. (F) The $R_{AH}$ and $R_{xx0}$ at different temperature when $V_{top} = -16$ V and $V_{back} = -80$ V.

**Figure 3. The comparison between the previous/reversed AHE.**

(A, B) The $\mu_0 H_{C1}$ and $\mu_0 H_{C2}$ extracted from figure 2. The $\mu_0 H_{C1}$ for negative AHE keeps nearly unchanged, while the $\mu_0 H_{C1}$ for positive AHE decreases with gate voltage decreasing. (C) The first-order differential of $R_{xx0}$ to gate voltages. The peak is related to the crossover from the negative AHE to positive AHE. (D) The relationship between $\sigma_{AH}$ and $\sigma_{xx}$. (E, F) The AHE in another 5 SLs MBT. The AHE at different gate voltages also shows a crossover from negative AHE to positive AHE (E). The $R_{xx0}$ mapping at different gate voltages is shown, and the maximum of the $R_{xx0}$ is corresponding to the max value of $R_{AH}$ (yellow dots), indicating by the yellow dash line (F).

**Figure 4. Modeling the AHE reversal by the competition between the Berry curvature and skew scattering.**

(A) The band structure of the 5SL MBT thin film. (B) The anomalous Hall conductivity as a function of the energy with the total magnetization $M_Z > 0$. (C) A schematic diagram of the negative AHE from the Berry curvature (top panel) and positive AHE from the skew scattering (bottom panel).

**Figure 1. The 5 SL MBT device and the onset of the reversed AHE.**

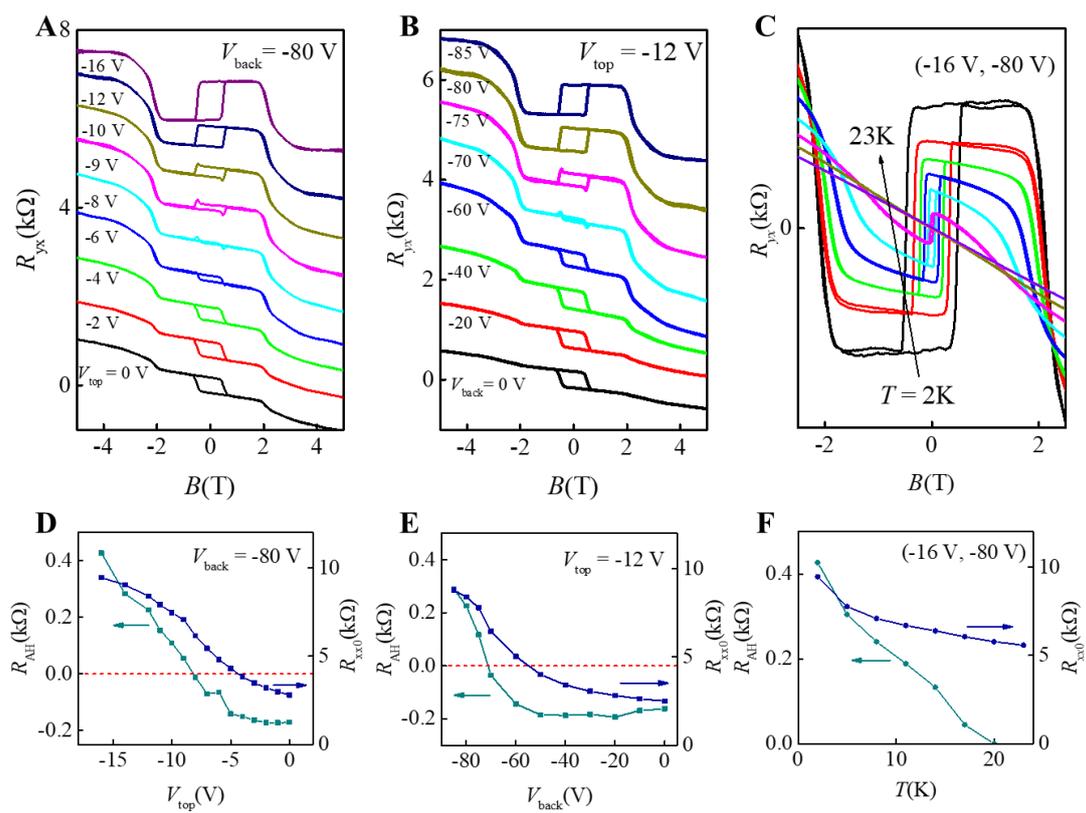

**Figure 2. The AHE sign reversal and its temperature dependence.**

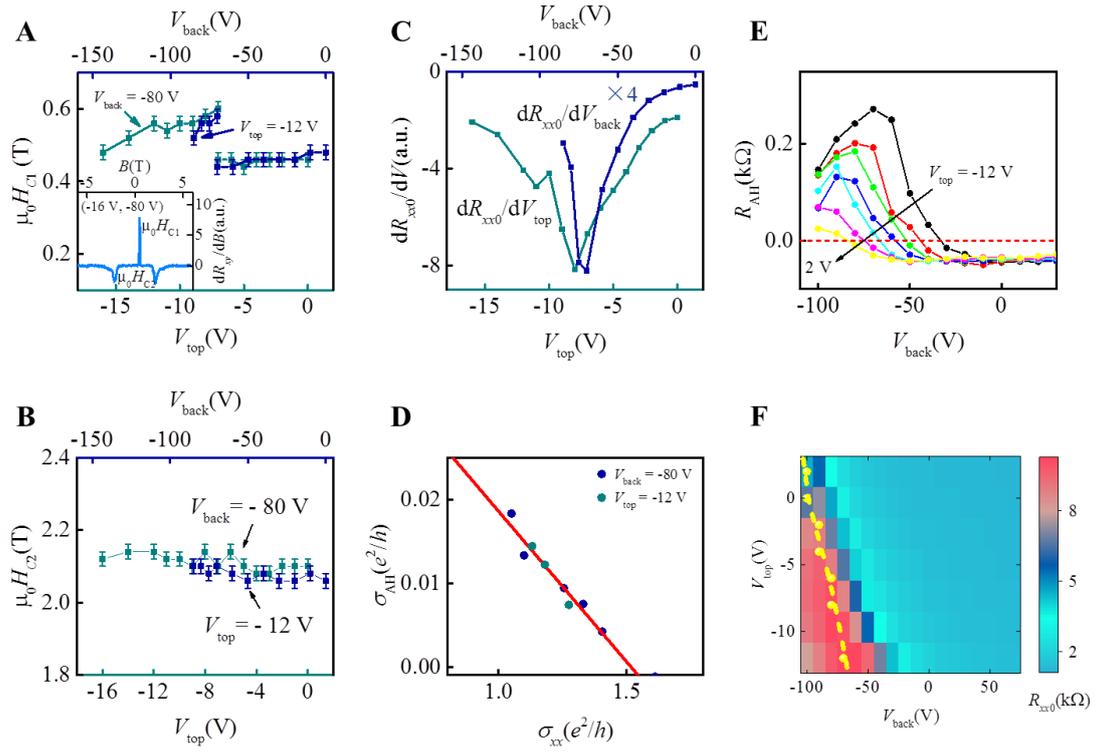

**Figure 3. The comparison between the previous/reversed AHE.**

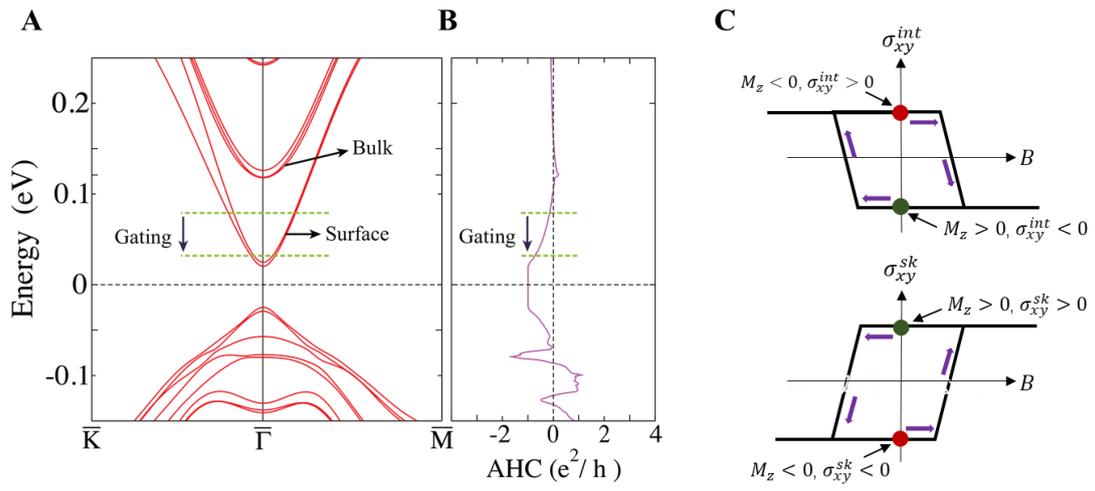

**Figure 4. Modeling the AHE reversal by the competition between the Berry curvature and skew scattering.**